\documentclass[11pt]{article}
\usepackage[T1]{fontenc}
\usepackage[utf8]{inputenc}
\usepackage[english]{babel}
\usepackage{geometry}
\usepackage{amsfonts}
\usepackage{amsmath}
\usepackage{amsthm}
\usepackage[title]{appendix}
\usepackage{amssymb}
\usepackage{tensor}
\usepackage{hyperref}
\usepackage[square, numbers]{natbib}
\usepackage{authblk}
\hypersetup{
    colorlinks,
    citecolor=red,
    linkcolor=blue,
    urlcolor=red
}

\title{A Nieh-Yan-like topological invariant in General Relativity}
\author[1]{G. Pollari \thanks{Electronic address: 504077@mail.muni.cz}}
	\affil[1]{Department of Theoretical Physics and Astrophysics,
		Faculty of Science of the Masaryk University,
		Kotlářská 2, 611 37 Brno, Czech Republic}
	
\begin{document}
	\maketitle
	\newcommand{\I}{\indices}
    \newcommand{\p}{^{(\phi)}}
	\renewcommand{\thesection}{\Roman{section}}
	\renewcommand{\thesubsection}{\thesection.\Roman{subsection}} 

\begin{center}
	\textbf{Abstract}
\end{center}

	In the present work we will introduce and prove a topological invariant term in General Relativity involving the torsion tensor that has never been showed before. Such a term is a slight modification of the Nieh-Yan four-form and likewise it stems from a Chern-Simons three-form. We provide the proof in both holonomic and orthogonal basis and show that its integral reduces to a boundary term that vanishes with the right conditions. As all topological invariant objects the new term does not affect the Einstein field equations in pure gravity, but when matter fields couple to the gravitational field, the torsion tensor arises and its contribution changes the "rules of the game". Therefore it is of great importance to study how those pieces irrelevant in bare gravity modify the interaction with fields of the Standard Model.

	\section{Introduction}
	The theory of General Relativity (GR) was proven to be the most accurate description of gravity \citep{Wald} \citep{Thorne}. There are two equivalent ways of writing the action of pure gravity: \textit{first} and \textit{second order} formalisms. The latter corresponds to the Einstein-Hilbert (EH) action whose connection depends on the metric tensor and is Levi-Civita, i.e. torsion free. The first is the Palatini (HP) action where the connection is considered to be independent from the metric. The variation of the HP action with respect to the connection leads to the torsion-free condition and both the EH and the P actions give the same Einstein field equations. However, the equations of motion remain unchanged also by adding topological invariants of the curvature and the torsion. The first  discovered was the Gauss-Bonnet-Chern \citep{Chern1},\citep{Chern2} giving the Euler characteristic of the manifold. Another topological piece not involved in the dynamics is the Pontryagin density which is related to the Chern-Simons current for GR \citep{Pontryagin1} \citep{Pontryagin2}. The last one that joined the family was found in 1982 by Nieh and Yan \citep{Nieh} which stemmed from the generalized Bianchi identities and then was expressed in terms of a Chern-Simons three-form \citep{Pontryagin2}. \\ \noindent
	For decades the quantization of gravity and its coupling with matter fields of the SM has been the scope of Quantum Gravity. Loop Quantum Gravity (LQG) is one of the most well-developed attempts constructed from GR that expresses gravity as a Yang-Mills SU(2) gauge theory \citep{Thiemann} \citep{Rovelli} \citep{Gambini} \citep{Gambini2} \citep{CQG}. In 1996 S\"{o}ren Holst proved that there exists another topological invariant in the Palatini formulation \citep{Holst} from which the Ashtekar-Barbero connection naturally arises from the action. This allows the GR action to be SU(2) invariant as a consequence of the presence of the Gauss constraint in the Hamiltonian picture. From such a Holst invariant the Immirzi parameter $\beta:= 1/\gamma$ emerges as an arbitrary constant, entering the value of the area operator spectrum \citep{Rovelli} \citep{Rovelli2} $\textbf{A}_{\vec{j}} = 8\pi \beta \hslash G c^{-3} \sum_{i} \sqrt{j_i\left(j_i+1\right)}$, therefore its meaning is crucial both in a topological and in a more pragmatic way. The first value obtained $\beta= ln2 /\pi \sqrt{3}$ was found by Ashtekar \textit{et al.} \citep{AshtekarImm} by comparing the classical Bekerstein-Hawking entropy of black holes to the quantum area of LQG. A further study for different types of black hole horizons suggested slight different values \citep{AshtekarImm2}. However, the debate around it still continues and it seems that the Immirzi parameter might not be fixed to remove any ambiguity. Another possibility taken into account in \citep{CM1} \citep{Mercuri1} was to promote the Immirzi parameter to a real scalar field both in the Holst and Nieh-Yan action and obtain the effective action which turns out to be a scalar-tensor of the Immirzi field minimally coupled to gravity. \\ \noindent
	The presence of these terms in pure gravity do not affect the equations of motion, however the addition of matter fields such as fermions makes the first and second order formulations differ \citep{Rovelli} \citep{Rovelli2}. Therefore all the topological invariants have to be looked at in a different light and can play a crucial role in the dynamics of the gravity-matter coupling. \\ \noindent
	As we will see in Section IV, the Nieh-Yan-like topological invariant form, unlike the original Nieh-Yan, cannot be written in terms of curvature and torsion in the first order formulation, but it emerges as a part of the spin curvature when the spin connection is split as a sum of the Lorentz one and the contorsion. \noindent	
	In the following we will use the convention under which the Levi-Civita connection $\bar{\Gamma} \I{_\mu _\nu ^\rho} = \Gamma \I{_{(\mu}_{\nu)}^\rho} $ is symmetric in the first two indices and the torsion $T \I{_\mu _\nu ^\rho} = 2\Gamma \I{_{[\mu}_{\nu]}^\rho}$ is skew-symmetric in the first two, while the contorsion tensor is antisymmetric in the last two indices $K \I{_\mu _\nu_\rho} = - K \I{_\mu _\rho_\nu}$. Greek letters $\mu,\nu,\ldots =0,1,2,3$ are used for spacetime indices and Latin $I,J,\ldots = 0,1,2,3$ for Lorentz (flat) ones. The signature is $(-,+,+,+)$. Since the metric tensor can be written in an orthonormal basis via the tetrads $g_{\mu \nu} = e_\mu^I e_\nu^J \eta_{IJ}$, in order to avoid confusion, we denote as $\nabla_\mu$ a $g$-compatible affine connection and with $D_\mu$ the connection compatible with the tetrads, i.e. $D_\mu e_\nu^I = \nabla_\mu e_\nu^I + \omega \I{_\mu^I_J} e_\nu^J =0$ where $\omega \I{_\mu^I_J}$ is the spin connection antisymmetric in the last two indices.
\\ \noindent
	In this paper we introduce and prove a new topological invariant in two different basis and compare it with the Nieh-Yan term pointing out the differences. In Section II we start examining the Nieh-Yan four-form in a coordinate basis and introduce the new related object. In Section III we prove its topological invariant nature in both holonomic and orthonormal basis. We see that it cannot be written in terms of the curvature and torsions in first order formalism unlike the Nieh-Yan term but it is part of the decomposition of the curvature. Moreover, we show how it changes under a conformal transformation and that the action in a theory with torsion preserves its form with the aforementioned term. In Section IV we remark the conclusions of our work stressing the future applications. In Appendix A we compute all the steps for proving our claim in the orthonormal base.
 \section{The new term}
	The Nieh-Yan 4-form given by
	\begin{equation}\label{eq:2.1}
	    d\left( e^I \wedge T_I \right) = T^I \wedge T_I - e_I \wedge e_J \wedge R \I{^I^J}
	\end{equation}
	by Stokes theorem reduces to a surface integral that vanishes as some component of the torsion does on the boundary. Some authors say that it is a topological invariant by imposing a vanishing torsion at the boundary, but doing so they implicitly claim that all the components have to equal zero. On the contrary, the boundary conditions to compel are weaker as we will show below. Writing \eqref{eq:2.1} in holonomic basis:
    $$ d\left( e^I \wedge T_I \right) = \dfrac{1}{2} d\left( e^I_\nu T \I{_\rho_\sigma_I} dx^\nu \wedge dx^\rho \wedge dx^\sigma \right) = \dfrac{1}{2} \left( e^I_\nu T \I{_\rho_\sigma_I} \right)_{,\mu}  dx^\mu \wedge dx^\nu \wedge dx^\rho \wedge dx^\sigma$$
    which leads to
    \begin{equation}\label{eq:2.2}
       d\left( e^I \wedge T_I \right) = \dfrac{1}{2}\left( e^I_\nu T \I{_\rho_\sigma_I} \right)_{,\mu} \epsilon\I{^\mu^\nu^\rho^\sigma} d^4x
    \end{equation}
    where, $ (T_I)_{\mu \nu}:= \dfrac{1}{2} T_{\mu \nu I} $ and  $\epsilon^{\mu \nu \rho \sigma}$ is the Levi-Civita symbol, related to the Levi-Civita tensor $\varepsilon_{\mu \nu \rho \sigma}$ by $\varepsilon_{\mu \nu \rho \sigma} = \sqrt{-g} \epsilon_{\mu \nu \rho \sigma}$ and $\varepsilon^{\mu \nu \rho \sigma} = \epsilon^{\mu \nu \rho \sigma} /\sqrt{-g}$. Thus, contracting the Lorentz indices and integrating over all the spacetime $\mathcal{M}$ we obtain
    \begin{equation}\label{eq:2.3}
        \int_\mathcal{M} d\left( e^I \wedge T_I \right) = \dfrac{1}{2}\int_\mathcal{M} d^4x \left(T \I{_\rho_\sigma_\nu} \right)_{,\mu} \epsilon\I{^\mu^\nu^\rho^\sigma}.
    \end{equation}
    Since the partial derivative of the Levi-Civita symbol vanishes, then
    \begin{equation}\label{eq:2.4}
        \int_\mathcal{M}  d\left( e^I \wedge T_I \right) = \dfrac{1}{2}\int_\mathcal{M} d^4x \partial_\mu \left(T \I{_\rho_\sigma_\nu} \epsilon\I{^\mu^\nu^\rho^\sigma} \right) = \dfrac{1}{2}\int_\mathcal{M} d^4x \partial_\mu \left( \sqrt{-g}T \I{_\rho_\sigma_\nu} \varepsilon\I{^\mu^\nu^\rho^\sigma} \right)
        \end{equation}
    which vanishes iff 
    $$\left(T \I{_\rho_\sigma_\nu} \varepsilon\I{^\mu^\nu^\rho^\sigma}\right)|_{\partial \mathcal{M}} =0 .$$
    It is useful to decompose the torsion tensor in its irreducible components according to the Lorentz group \citep{Shapiro} \citep{CM1} \citep{Capozziello}:
    \begin{equation}\label{eq:2.5}
        T_{\mu \nu \rho} = \dfrac{1}{3} \left( T_\nu g_{\mu \rho} - T_\mu g_{\nu \rho} \right) - \dfrac{1}{6} \varepsilon \I{_\mu_\nu_\rho_\sigma} S^\sigma + q \I{_\mu_\nu_\rho}
    \end{equation}
    where $T^\mu = T \I{^\nu^\mu_\nu}$ is called the \textit{trace vector} and carries 4 d.o.f., $S^\sigma$ is called the \textit{(pseudotrace) axial vector} with 4 d.o.f. and $q \I{_\mu_\nu_\rho}$ is the non totally skew-symmetric traceless part of the torsion that has the remaining 16 d.o.f. and such that $\epsilon \I{^\mu^\nu^\rho^\sigma} q_{\mu \nu \rho} =0$. \\ \noindent
    Inserting \eqref{eq:2.5} into \eqref{eq:2.4} and contracting, it gives
    \begin{equation}\label{eq:2.6}
        \int_\mathcal{M} d\left( e^I \wedge T_I \right) = \dfrac{1}{2} \int_\mathcal{M} d^4x \sqrt{-g} \bar{\nabla}_\mu S^\mu
        \end{equation}
    which agrees with \citep{CM1}. Therefore only the axial vector has to vanish on the boundary and not all the other irreducible components. Following the same concepts, if 
    $$ \int_\mathcal{M} d\left( e^I \wedge T_I \right) $$
    by Stokes theorem vanishes because all or some of the components of $T_I$ do on the boundary, so has to do the four form
    \begin{equation}\label{eq:2.7}
    d\left(e^I \wedge \star T_I\right)
    \end{equation}
    where $\star$ is the Hodge operator. Indeed \eqref{eq:2.7} is the only other non trivial exact (so closed) four form that can be constructed from the torsion. The Hodge operator changes the components of the torsion two form via the Levi-Civita symbol but if we impose that $T_{\mu \nu \rho}$ vanishes, also the "hodged" one has to. Hence, the new term we want to study in the next section is \eqref{eq:2.7}.

\section{Direct evaluation}
\subsection{Holonomic basis}
 Expressing the torsion 2-form in holonomic basis and then apply the Hodge operator we have
	$$\star T_I =\dfrac{1}{2} T\I{_\alpha_\beta_I} \star \left(dx^\alpha \wedge dx^\beta \right) = \dfrac{1}{4} \sqrt{-g} T\I{_\alpha_\beta_I} g^{\alpha \gamma} g^{\beta \delta} \epsilon\I{_\gamma_\delta_\rho_\sigma} dx^\rho \wedge dx^\sigma $$
    which put into \eqref{eq:2.7} gives
    \begin{equation}\label{eq:3.1}
    d\left(e^I \wedge \star T_I\right) = \dfrac{1}{4} \left(\sqrt{-g} e_\nu^I T\I{_\alpha_\beta_I} g^{\alpha \gamma} g^{\beta \delta} \epsilon\I{_\gamma_\delta_\rho_\sigma}\right)_{,\mu} dx^\mu \wedge dx^\nu \wedge dx^\rho \wedge dx^\sigma =\dfrac{1}{4} \left( \sqrt{-g} T\I{_\alpha_\beta_\nu} g^{\alpha \gamma} g^{\beta \delta} \epsilon\I{_\gamma_\delta_\rho_\sigma}\right)_{,\mu} \epsilon^{\mu \nu \rho \sigma} d^4x.
    \end{equation}
   Thus, eq.\eqref{eq:3.1} becomes
    $$d\left(e^I \wedge \star T_I\right) = \dfrac{1}{4}  \left( \sqrt{-g} T\I{^\gamma^\delta_\nu} \epsilon\I{_\gamma_\delta_\rho_\sigma}\right)_{,\mu} \epsilon^{\mu \nu \rho \sigma} d^4x = \dfrac{1}{4} \sqrt{-g} \: \left(\bar{\nabla}_\mu T\I{^\gamma^\delta_\nu}  \epsilon\I{_\gamma_\delta_\rho_\sigma}\epsilon^{\mu \nu \rho \sigma} \right) d^4x.$$
    Contracting all the quantities and using \eqref{eq:2.5} we obtain
    \begin{equation}\label{eq:3.2}
    	d\left(e^I \wedge \star T_I\right) = \sqrt{-g} \: \bar{\nabla}_\mu T\I{^\mu} d^4x.
    \end{equation}
    Therefore, its integral
    \begin{equation}\label{eq:3.3}
    	\int_\mathcal{M} d\left(e^I \wedge \star T_I\right) = \int_\mathcal{M} d^4x \;\sqrt{-g} \, \bar{\nabla}_\mu T\I{^\mu}= \int_\mathcal{M} d^4x \; \partial_\mu \left(\sqrt{-g} \, T\I{^\mu} \right)
    \end{equation}
    by the divergence theorem reduces to a surface integral that vanishes iff $T\I{^\mu}|_\mathcal{M}=0$. It is striking that in the Nieh-Yan case the totally antisymmetric part of the torsion, i.e. the axial vector, has to vanish whereas in the new term its trace part.
    \subsection{Orthonormal basis}
    Expressing the torsion form in orthonormal basis $\{e^I\}$:
    $$\star T_I = \dfrac{1}{4} T \I{_\mu_\nu_I} e^\mu_L e^\nu_K \epsilon \I{^L^K_J_M} e^J \wedge e^M = \left(C_I\right) \I{_J_M} e^J \wedge e^M$$
    where we set
    $$\left(C_I\right) \I{_J_M} = \dfrac{1}{4} T\I{_\mu_\nu_I} e^\mu_R e^\nu_S \epsilon\I{^R^S_J_M}$$
    the three form in brackets in \eqref{eq:2.7} becomes
	$$ e^I \wedge \star T_I = \left(C_I\right) \I{_J_M} e^I \wedge e^J \wedge e^M$$
	and its differential
	$$ d\left(e^I \wedge \star T_I\right) = \left(C_I\right) \I{_J_M_{,N}} e^N \wedge e^I \wedge e^J \wedge e^M + \left(C_I\right) \I{_J_M} de^I \wedge e^J \wedge e^M - \left(C_I\right) \I{_J_M} e^I \wedge de^J \wedge e^M + \left(C_I\right) \I{_J_M} e^I \wedge e^J \wedge de^M$$
	$$ = \left(C_I\right) \I{_J_M_{,N}} e^N \wedge e^I \wedge e^J \wedge e^M + \left(C_I\right) \I{_J_M} de^I \wedge e^J \wedge e^M - \left(C_I\right) \I{_J_M} e^I \wedge e^M \wedge de^J + \left(C_I\right) \I{_J_M} e^I \wedge e^J \wedge de^M$$
	\begin{equation}\label{eq:3.4}
		d\left(e^I \wedge \star T_I\right) = \left(C_I\right) \I{_J_M_{,N}} e^N \wedge e^I \wedge e^J \wedge e^M + \left(C_I\right) \I{_L_M} de^I \wedge e^L \wedge e^M + 2\left(C_I\right) \I{_L_M} e^I \wedge e^L \wedge de^M
	\end{equation} 
	where we used the antisymmetry of $\left(C_I\right) \I{_J_M}$ under $J \leftrightarrow M$. \\ \noindent 
	By the Cartan structure equations :
    \begin{equation}\label{eq:3.5}
	    R^{IJ} = d\omega\I{^I^J} + \omega \I{^I_K} \omega \I{^K^J}
	\end{equation}
	\begin{equation}\label{eq:3.6}
	    T^I = de^I + \omega \I{^I_J} \wedge e^J,
	\end{equation}
	from \eqref{eq:3.6} we relate the torsion and the contorsion tensors:
    $$T \I{_\mu_\nu_\rho} = \left(K \I{_\mu_\nu_\rho} - K \I{_\nu_\mu_\rho} \right)$$
    where the contorsion in terms of the torsion is given by
    \begin{equation}\label{eq:3.7}
        K \I{_\mu_\nu_\rho} = \dfrac{1}{2} \left( T \I{_\mu_\nu_\rho} + T \I{_\rho_\nu_\mu} + T\I{_\rho_\mu_\nu}\right)
    \end{equation}
    so that $\Gamma \I{_\mu_\nu^\rho} = \bar{\Gamma} \I{_\mu_\nu^\rho} + K \I{_\mu_\nu^\rho}$, being $\bar{\Gamma} \I{_\mu_\nu^\rho}$ the Christoffel symbol. Using \eqref{eq:3.6} in \eqref{eq:3.4} it yields
	$$ d\left(e^I \wedge \star T_I\right) = \left(C_I\right) \I{_J_M_{,N}} e^N \wedge e^I \wedge e^J \wedge e^M +\left(C_I\right) \I{_L_M} T^I \wedge e^L \wedge e^M - \left(C_I\right) \I{_N_M} \omega \I{^I_J} \wedge e^J \wedge e^N \wedge e^M $$
	$$+ \left(C_I\right) \I{_N_M}  e^I \wedge e^N \wedge T \I{^M} - 2\left(C_I\right) \I{_N_M}  e^I \wedge e^N \wedge \omega \I{^M_J}  \wedge e^J$$
	$$ = - \left(C_I\right) \I{_J_M_{,N}} e^I \wedge e^J \wedge e^M \wedge e^N +  \dfrac{1}{2}\left(C_I\right) \I{_L_M} T\I{_\mu_\nu^I} e^\mu_J e^\nu_N e^J \wedge e^N \wedge e^L \wedge e^M - \left(C_I\right) \I{_N_M} \omega \I{_\mu^I_J} e^\mu_L e^L \wedge e^J \wedge e^N \wedge e^M $$
	$$+\left(C_I\right) \I{_N_M} T\I{_\mu_\nu^M} e^\mu_J e^\nu_L e^I \wedge e^N \wedge e^J \wedge e^L - 2 \left(C_I\right) \I{_N_M} \omega \I{_\mu^M_J} e^\mu_L e^I \wedge e^N \wedge e^L \wedge e^J$$
	Setting
	\begin{subequations}
		\begin{align}\label{eq:3.8}
			&\left(C_I\right) \I{_J_M} =  \dfrac{1}{4} T\I{_\mu_\nu_I} e^\mu_R e^\nu_S \epsilon\I{^R^S_J_M} \\
			&\left(B^L\right)\I{_J_N} = \dfrac{1}{2} T\I{_\mu_\nu^L}e^\mu_J e^\nu_N\\
			&\left(A\I{^L_N}\right)_J = \omega\I{_\mu^L_N} e^\mu_J
		\end{align}
	\end{subequations}
	and arranging the tetrad indices we obtain:
	$$	d\left(e^I \wedge \star T_I\right) = \Big[ - \left(C_I\right)\I{_J_M_{,N}} - \left(C_L\right)\I{_I_M} \left(B^L\right) \I{_J_N} + \left(C_L\right)\I{_N_M} \left(A\I{^L_J}\right)\I{_I} $$
    $$+ 2\left(C_I\right)\I{_N_L} \left(B^L\right)\I{_J_M} + 2 \left(C_I\right)\I{_N_L} \left(A \I{^L_J}\right)_M \Big] e^I \wedge e^J \wedge e^M \wedge e^N$$
leading to
		$$d\left(e^I \wedge \star T_I\right) = e \Big[ - \left(C_I\right)\I{_J_M_{,N}} - \left(C_L\right)\I{_I_M} \left(B^L\right) \I{_J_N} + \left(C_L\right)\I{_N_M} \left(A\I{^L_J}\right)\I{_I} $$ 
    \begin{equation}\label{eq:3.9}
    + 2 \left(C_I\right)\I{_N_L} \left(B^L\right)\I{_J_M} + 2 \left(C_I\right)\I{_N_L} \left(A \I{^L_J}\right)_M \Big] \epsilon\I{^I^J^M^N} d^4x
	\end{equation}
    where we made use of the formula $e^I \wedge e^J \wedge e^M \wedge e^N = e\epsilon\I{^I^J^M^N} d^4x$ with $e := det\left( e_\mu^I\right) = \sqrt{-g}$.
    In Appendix A we calculated all the terms of \eqref{eq:3.9}. 
    Inserting \eqref{eq:A.1}, \eqref{eq:A.2}, \eqref{eq:A.3}, \eqref{eq:A.6}, \eqref{eq:A.7}  into \eqref{eq:3.9}, it yields
		$$d\left(e^I \wedge \star T_I\right) = e \Bigl[ \nabla_\mu T\I{_\nu^\mu^\nu} -  T \I{_\mu_\nu^\mu} \omega \I{_\rho^\rho^\nu} - \dfrac{1}{2} T \I{_\mu_\nu_\rho} T \I{^\mu^\nu^\rho} + T \I{^\mu_\nu_\rho} \omega \I{_\mu^\rho^\nu} -  T \I{_\mu_\nu^\mu} T \I{_\rho^\nu^\rho} $$
  $$+ \dfrac{1}{2} T \I{_\mu_\nu_\rho} T \I{^\mu^\nu^\rho} + T \I{_\mu_\nu^\mu} \omega \I{_\rho^\rho^\nu}  + T \I{_\mu^\nu_\rho} \omega \I{_\nu^\rho^\mu} \Bigr] d^4x$$
  and by exploiting the antisymmetry of the torsion tensor we obtain
  $$ d\left(e^I \wedge \star T_I\right)= e \left[ \nabla_\mu T\I{_\nu^\mu^\nu} - T \I{_\mu_\nu^\mu} T \I{_\rho^\nu^\rho} \right]d^4x.$$
  Decomposing the connection into its symmetric and antisymmetric part, using the relation \eqref{eq:3.7} and \eqref{eq:2.5} we finally conclude that
  \begin{equation}\label{eq:11}
      d\left(e^I \wedge \star T_I\right)= e\bar{\nabla}_\mu T\I{^\mu}d^4x
  \end{equation}
  coinciding with \eqref{eq:3.2} as claimed.
  
  \subsection{Decomposition}
  The RHS of \eqref{eq:2.1} is equivalent to its LHS by using the Cartan structure equations \eqref{eq:3.5}, \eqref{eq:3.6}. We would like to write, if possible, the four form \eqref{eq:2.7} in first order formalism. However, unlike the other topological invariants covered in literature, such a term cannot be a combination of the torsion and the curvature maintaining the connection as independent. First, we write the action
  \begin{equation}\label{eq:4.0}
      S = \int_\mathcal{M} d^4x \; e \: e^\mu_I e^\nu_J R \I{_\mu_\nu^I^J}
  \end{equation}
  decompose the spin connection $\omega \I{_\mu^I^J} = \bar{\omega} \I{_\mu^I^J} - K \I{_\mu^I^J}$:
  \begin{equation}\label{eq:4.1}
  	 S = \int_\mathcal{M} d^4x \; e \,\left[ e^\mu_I e^\nu_J \bar{R} \I{_\mu_\nu^I^J} - 2 e^\mu_I e^\nu_J \bar{D}_\mu K \I{_\nu^I^J} + e^\mu_I e^\nu_J \left( K \I{_\mu^I^N} K \I{_\nu_N^J} - K \I{_\nu^I^N} K \I{_\mu_N^J} \right)  \right]
  \end{equation}
  where $\bar{D}_\mu$ is the Levi-Civita tetrad-compatible connection. Contracting the second term of \eqref{eq:4.1} and using \eqref{eq:2.5} and \eqref{eq:3.7} we end up with
  \begin{equation}\label{eq:4.2}
  	 S = \int_\mathcal{M} d^4x \; e \,\left[ e^\mu_I e^\nu_J \bar{R} \I{_\mu_\nu^I^J} - 2 \bar{\nabla}_\mu T^\mu + \left( K \I{_\mu^\mu^\rho} K \I{_\nu_\rho^\nu} - K \I{_\nu^\mu^\rho} K \I{_\mu_\rho^\nu} \right)  \right].
  \end{equation}
  From \eqref{eq:4.2} we see that the integral of \eqref{eq:2.7} corresponds to the Levi-Civita covariant derivative of the trace vector coming from the decomposition of the connection. This is not surprising whatsoever. The integral of the Nieh-Yan form \eqref{eq:2.1}, following the same procedure we have just shown, leads to \eqref{eq:2.6}. In particular, the contorsion terms of the splitting of the connection in the curvature cancel the torsion-torsion part and the curvature constructed from the Levi-Civita connection vanishes by virtue of the Bianchi identity. Therefore, in that case the only surviving term is exactly \eqref{eq:2.6}. However, in \eqref{eq:4.2} the presence of the curvature does not allow to recast $d(e^I \wedge \star T_I)$ in a way in which we have a curvature with a connection independent from the metric. From \eqref{eq:4.2} we can single out  the topological term and write
  \begin{equation}\label{eq:4.3}
  	  2d(e^I \wedge \star T_I) = \star \left(e^I \wedge e^J\right) \wedge \left( \bar{R}_{JI} - {R}_{JI} + (K \I{_I^K} \wedge K \I{_K_J}) \right)= \star \left(e^I \wedge e^J\right) \wedge \bar{D}K_{IJ}
  \end{equation}
  where
  $$ \bar{D}K_{IJ} = dK_{IJ} + 2\bar{\omega}\I{_{[I|}^K} \wedge K \I{_K_{|J]}}$$
  is the Levi-Civita covariant derivative of the contorsion with respect to a non-coordinate basis. Equivalently, the action \eqref{eq:4.0} can also be rewritten in terms of forms as 
  \begin{equation}\label{eq:4.01}
  	  S = \int_{\mathcal{M}} e^I \wedge e^J \wedge \star R_{JI} = \int_{\mathcal{M}} \left[\star \left( e^I \wedge e^J\right) \wedge \left(R_{JI} + K \I{_I^K} \wedge K \I{_K_J}\right) - 2 d(e^I \wedge \star T_I) \right]
  \end{equation}
  Clearly \eqref{eq:2.7} is Diff($\mathcal{M}$) and SO(1,3) invariant. We would like to know how it transforms under a conformal transformation and how \eqref{eq:4.0} changes. In the Cartan structure equation \eqref{eq:3.6} the spin connection is a gauge field, treated as an independent variable, unchanged under any transformation of the metric. In accordance with \citep{Shapiro}, under the conformal transformation 
  \begin{equation}\label{eq:4.4}
      \tilde{g}_{\mu \nu}(x) = \phi^2(x) g_{\mu \nu}(x) \hspace{2cm} \tilde{e}^I(x) = \phi(x) e^I(x) ,
  \end{equation}
  with $\phi \in \Omega^0(\mathcal{M})$, eq.\eqref{eq:3.6} becomes
    $$ \tilde{T}^I = d\tilde{e}^I + \omega \I{^I_J} \wedge \tilde{e}^J$$
    from which we obtain
    \begin{equation}\label{eq:4.5}
        \tilde{T}^I = \phi T^I + d\phi \wedge e^I.
    \end{equation}
  It is easy to see that by expressing \eqref{eq:4.5} in components we have
  $$ \tilde{T} \I{_\mu_\nu^\rho} = \tilde{T} \I{_\mu_\nu^I} \tilde{e}^\rho_I = T \I{_\mu_\nu^\rho} + \dfrac{1}{\phi}\left( \phi_{,\mu} \delta_\nu^\rho - \phi_{,\nu} \delta_\mu^\rho \right)$$
  and by using the decomposition \eqref{eq:2.5}
  the trace vector changes as
  \begin{equation}\label{eq:4.6}
      \tilde{T}_\mu = T_\mu - \dfrac{3}{\phi}\phi_{,\mu}.
  \end{equation}
  By virtue of this we can write the conformally transformed term \eqref{eq:2.7} as
  $$ d(\tilde{e}^I \wedge \star \tilde{T}_I) = d(\phi e^I \wedge \star ( \phi T_I + d\phi \wedge e_I))$$
  $$ = \dfrac{1}{4} d \left(\sqrt{-g} \phi e_\nu^I g^{\alpha \gamma} g^{\beta \delta} \varepsilon_{\alpha \beta \rho \sigma} \left(\phi T_{\gamma \delta I} + (\phi_{,\gamma} e_{\delta I} - \phi_{,\delta} e_{\gamma I}) \right) \right) dx^\nu \wedge dx^\rho \wedge dx^\sigma$$
    $$ -\dfrac{1}{2} \left(\sqrt{-g} \phi g^{\alpha \gamma} g^{\beta \delta} \left( \phi T_{\gamma \delta \nu} + \phi_{,\gamma} g_{\delta \nu} - \phi_{,\delta} g_{\gamma \nu}\right)\right)_{,\mu} (\delta^\mu_\alpha \delta^\nu_\beta - \delta^\mu_\beta \delta^\nu_\alpha)  d^4x,$$
    using \eqref{eq:2.5} and contracting it yields
    \begin{equation}\label{eq:4.7}
        d(\tilde{e}^I \wedge \star \tilde{T}_I) = \sqrt{-g} \bar{\nabla}_\mu \left( \phi^2 T^\mu - 3\phi \phi \I{_,^\mu} \right) d^4x = \sqrt{-g} \bar{\nabla}_\mu (\phi^2 \tilde{T}^\mu) d^4x
    \end{equation}
    where in the last line we used \eqref{eq:4.6}. It is not surprising that \eqref{eq:4.7} is again a boundary term since \eqref{eq:2.7}, by Stokes theorem, reduces to a surface integral independently from the basis chosen. Now we want to derive \eqref{eq:4.0} under a conformal transformation \eqref{eq:4.4} in presence of torsion. We denote the torsion-free $\tilde{e}$-compatible connection as 
    \begin{equation}\label{eq:4.8}
        \tilde{D}_\mu \tilde{e}_\nu^I = \tilde{e}^I_{\nu,\mu} - \tilde{\Gamma} \I{_\mu_\nu^\rho} \tilde{e}^I_\rho + \tilde{\omega} \I{_\mu^I_J} \tilde{e}^J_\nu =0
    \end{equation}
    where $\tilde{\Gamma} \I{_\mu_\nu^\rho} = \bar{\Gamma} \I{_\mu_\nu^\rho} + \tensor{\p\Gamma}{_\mu_\nu^\rho}$ and $\tilde{\omega} \I{_\mu^I^J} = \bar{\omega} \I{_\mu^I^J} + \tensor{\p\omega}{_\mu^I^J}$, $\tensor{\p\Gamma}{_\mu_\nu^\rho}$ and $\tensor{\p\omega}{_\mu^I^J}$ being the part of the connection containing the derivative of $\phi$. The conformally transformed action \eqref{eq:4.0} is
    $$  \tilde{S} = \int_\mathcal{M} d^4x\; \tilde{e} \; \tilde{e}^\mu_I \tilde{e}^\nu_J \tilde{R} \I{_\mu_\nu^I^J} $$
    $$= \int_\mathcal{M} d^4x\; e \; \phi^2 e^\mu_I e^\nu_J \left[ 2\tilde{\omega}_{[\nu,\mu]} - 2\tilde{K}_{[\nu,\mu]} + 2 \left(\tilde{\omega} \I{_{[\mu}^I^K} - \tilde{K} \I{_{[\mu}^I^K}) \right) \left(\tilde{\omega} \I{_{\nu]}_K^J} - \tilde{K} \I{_{\nu]}_K^J}) \right) \right]$$
    where $\tilde{K} \I{_\nu^I^J}$ is the transformed contorsion tensor. Rearranging the terms it yields
    \begin{equation}\label{eq:4.9}
        \tilde{S} = \int_\mathcal{M} d^4x\; e \; \phi^2 e^\mu_I e^\nu_J \left( \tensor{\p\tilde{R}}{_\mu_\nu^I^J} - 2 \tilde{D}_{\mu} \tilde{K} \I{_{\nu}^I^J} + 2 \tilde{K} \I{_{[\mu}^I^K} \tilde{K} \I{_{\nu]}_K^J} \right)
    \end{equation} 
    with
    $$ \tensor{\p\tilde{R}}{_\mu_\nu^I^J} = 2\tilde{\omega}_{[\nu,\mu]} +2 \tilde{\omega} \I{_{[\mu}^I^K} \tilde{\omega} \I{_{\nu]}_K^J}.$$
    Using the compatibility of the connection \eqref{eq:4.8} and that the trace of the contorsion equals \eqref{eq:4.6}, the second term in \eqref{eq:4.9} gives
    $$ -2 \phi^2 e^\mu_I e^\nu_J \tilde{D}_{\mu} \tilde{K} \I{_{\nu}^I^J} = -2 \phi^2 \tilde{\nabla}_{\mu} \tilde{T}^\mu +4 \phi \phi_{,\mu} \tilde{T}^\mu, $$
    expressing the connection $\tilde{\nabla}_\mu$ on the trace vector in terms of $\bar{\nabla}_\mu$ it yields
    $$ \tilde{\nabla}_\mu \tilde{T}^\mu = \bar{\nabla}_\mu \tilde{T}^\mu + \dfrac{4}{\phi} \phi_{,\mu}\tilde{T}^\nu.$$
    Therefore the action \eqref{eq:4.9} becomes 
    \begin{equation}\label{eq:4.10}
        \tilde{S} = \int_\mathcal{M} d^4x\; e \; \left(\phi^2 e^\mu_I e^\nu_J \tensor{\p\tilde{R}}{_\mu_\nu^I^J}- 2 \bar{\nabla}_\mu \big(\phi^2 \tilde{T}^\mu\big) + 2 \phi^2 e^\mu_I e^\nu_J\tilde{K} \I{_{[\mu}^I^K} \tilde{K} \I{_{\nu]}_K^J} \right).
    \end{equation}
    We see that the second term in brackets of \eqref{eq:4.10} coincide (up to a factor 2) with \eqref{eq:4.7}. In our case we used the decomposition of the torsion tensor in its irreducible components but the equality mentioned holds also for the general undecomposed one. Hence, this allows to write the conformally transformed action in the language of forms as
    \begin{equation}\label{eq:4.11}
        \tilde{S} = \int_{\mathcal{M}} \tilde{e}^I \wedge \tilde{e}^J \wedge \star \tilde{R}_{JI} = \int_{\mathcal{M}} \left[\star \left( \tilde{e}^I \wedge \tilde{e}^J\right) \wedge \left(\tensor{\p\tilde{R}}{_J_I} + \tilde{K} \I{_I^K} \wedge \tilde{K} \I{_K_J}\right) - 2 d(\tilde{e}^I \wedge \star \tilde{T}_I) \right]
    \end{equation}
    thus, it is the same as just transforming every quantity in the action \eqref{eq:4.01} under \eqref{eq:4.4} while preserving its form.
  \section{Conclusions}
  In this paper we introduced a new topological invariant object in General Relativity driven by the isomorphism nature of the Hodge star applied to the torsion in the Nieh-Yan four form. We concluded that such a new term reduces to a covariant divergence over the entire manifold which vanishes as the trace vector does on the boundary. It is worth noting that the Nieh-Yan invariant does not affect the equation of motion as long as the totally antisymmetric part of the tensor irreducible components, i.e its pseudotrace axial vector, vanishes on the surface of the spacetime, while the new term requires that the trace vector does. Moreover, the Nieh-Yan topological invariant was first proven in its tensor nature, namely starting from the generalized Bianchi identities (also called Bianchi-Penrose symmetry) in 1982 and then compactly written as a differential of a Chern-Simons three form \eqref{eq:2.1}. Conversely, we proved the new topological invariant quantity beginning from its differential form in \eqref{eq:2.7}. Then we pointed out that such an invariant is hidden in the spin curvature of the GR action and arises when we split the connection into the Lorentz spin connection plus contorsion. As a consequence, the Nieh-Yan-like term cannot be recasted differently from expressing the metricity of the connection.  A work in which all the known topological invariants, except for the one we found, were analyzed in \citep{Perez} in the LQG framework. The implementation of \eqref{eq:2.7} into it, leading to the full action 
  $$ S = \int \Bigl( \underbrace{e^I \wedge e^J \wedge \star R_{JI}}_{HP} + \underbrace{\alpha_1 e^I \wedge e^J \wedge R_{IJ}}_{HOLST} + \underbrace{\alpha_2 R^{IJ} \wedge R_{JI}}_{PONTRYAGIN} + \underbrace{\alpha_3 R^{IJ} \wedge \star R_{JI}}_{GAUSS-BONNET} + \underbrace{\alpha_4 d(e^I \wedge T_I)}_{NIEH-YAN} + \underbrace{\alpha_5 d(e^I \wedge \star T_I)}_{NEW \: TERM} \Bigr)$$
    where $\{\alpha_i\}_{i=1}^5$ are arbitrary dimensionless constants, should be taken into account. Moreover, following \citep{CM1}, where the Barbero-Immirzi parameter was promoted to a real scalar field in the Holst and Nieh-Yan action, the addition of \eqref{eq:2.7} multiplied with a different scalar field generates a new independent massless kinetic term which couples minimally with gravity and potentially with other fields. A conformal transformation of the object we focus on in this work still preserves its topological nature and the form of the action \eqref{eq:4.11} makes us conclude that, when expressed in terms of \eqref{eq:2.7}, it is the same independently from the transformed metric. As a consequence, such a total derivative of the modified trace vector vanishes nonetheless.However, if we non-minimally attach a further scalar field $\psi$ to gravity, such as in a Brans-Dicke theory with torsion, then such a piece is no longer irrelevant and we will get an effective action with non standard kinetic terms of the coupled fields.
\section*{Acknowledgments} 
I would like to thank Professor Franz Hinterleitner of Masayk University for precious discussions and consultations I had with him. This work was supported by a Masaryk University Grant.
  \newpage
  \begin{appendices}
  \setcounter{equation}{0}
  \renewcommand{\theequation}{\thesection.\arabic{equation}}
  \section{}
\begin{itemize}
    \item TERM 1
    $$ \left(C_I\right)\I{_J_M_{,N}} \epsilon\I{^I^J^M^N} = \dfrac{1}{4} \left(T\I{_\mu_\nu_I} e^\mu_R e^\nu_S \epsilon \I{^R^S_J_M}\right)_{,\rho} e^\rho_N \epsilon \I{^I^J^M^N} $$
    since the Levi Civita symbol in brackets can be brought out and the only non contracted indices are the flat ones, the partial derivative can be replaced by a generic, possibly non torsion-free, covariant derivative
    $$ \dfrac{1}{4} e^\rho_N  \nabla_\rho \left(T\I{_\mu_\nu_I} e^\mu_R e^\nu_S \right) \epsilon \I{^R^S_J_M} \epsilon \I{^I^J^M^N} = -\dfrac{1}{2} e^\rho_N  \nabla_\rho \left(T\I{_\mu_\nu_I} e^\mu_R e^\nu_S \right) \left( \eta^{RI} \eta^{SN} - \eta^{RN} \eta^{SI}\right)$$
    $$- \dfrac{1}{2} e^\rho_N  \nabla_\rho \left(T\I{_\mu_\nu_I} e^{\mu I} e^{\nu N} - \dfrac{1}{2}T\I{_\mu_\nu_I} e^{\mu N} e^{\nu I} \right) = - \dfrac{1}{2}\nabla_\rho \left[ e^\rho_N T\I{_\mu_\nu_I} \left(e^{\mu I} e^{\nu N} - e^{\mu N} e^{\nu I} \right) \right] $$
    $$+ \dfrac{1}{2} \nabla_\rho \left(e^\rho_N\right) T\I{_\mu_\nu_I} \left(e^{\mu I} e^{\nu N} - e^{\mu N} e^{\nu I} \right).$$
    In order to simplify the RHS of the last line we make use of the relation between the covariant derivative compatible with the tetrads and the one with the metric tensor:
    $$D_\rho e^\rho_N = \nabla_\rho e^\rho_N - \omega \I{^L_N} e^\rho_L =0$$
    thus, contracting all the indices we obtain
    \begin{equation} \label{eq:A.1} \left(C_I\right)\I{_J_M_{,N}} \epsilon\I{^I^J^M^N} = -  \nabla_\mu T\I{_\nu^\mu^\nu} +  T \I{_\mu_\nu^\mu} \omega \I{_\rho^\rho^\nu}
    \end{equation}
    where
    $$\omega \I{_\rho^\rho^\nu} = e^\rho_I e^\nu_J \omega\I{_\rho^I^J} = e^\rho_I e^\nu_J \left( \bar{\omega}\I{_\rho^I^J} - K\I{_\rho^I^J}\right)$$
    with $\bar{\omega}\I{_\rho^I^J}$ and $ K\I{_\rho^I^J}$ the Levi Civita and the contorsion spin connection, respectively. \\
    \item TERM 2. \\
    $$\left(C_L\right)\I{_I_M} \left(B^L\right) \I{_J_N} \epsilon\I{^I^J^M^N} = - \dfrac{1}{8} T \I{_\mu_\nu_L} e^\mu_R e^\nu_S \epsilon \I{^R^S_I_M} \epsilon \I{^J^N_I_M} T \I{_\rho_\sigma^L} e^\rho_J e^\sigma_N$$
    $$ = \dfrac{1}{4}T \I{_\mu_\nu_L} T \I{_\rho_\sigma^L} e^\mu_R e^\nu_S e^\rho_J e^\sigma_N \left( \eta^{RJ} \eta^{SN} - \eta^{RN} \eta^{SJ} \right)= \dfrac{1}{4} T \I{_\mu_\nu_L} T \I{_\rho_\sigma^L} \left( g^{\mu \rho} g^{\nu \sigma} - g^{\mu \sigma} g^{\nu \rho} \right) $$
    giving 
    \begin{equation}\label{eq:A.2}
        \left(C_L\right)\I{_I_M} \left(B^L\right) \I{_J_N} \epsilon\I{^I^J^M^N} = \dfrac{1}{2} T \I{_\mu_\nu_\rho} T \I{^\mu^\nu^\rho}.
    \end{equation} \\
    \item TERM 3. \\
     $$\left(C_L\right)\I{_N_M} \left(A\I{^L_J}\right)\I{_I} \epsilon\I{^I^J^M^N} = - \dfrac{1}{4} T \I{_\mu_\nu_L} e^\mu_R e^\nu_S \epsilon \I{^R^S_N_M} \epsilon\I{^I^J^M^N} \omega \I{^L_J} e^\rho_I $$
     $$ = \dfrac{1}{2} T \I{_\mu_\nu_L} \omega \I{^L_J} \left( g^{\mu \rho} e^{\nu J} - g^{\nu \rho} e^{\mu J} \right)$$
     yielding
     \begin{equation}\label{eq:A.3}
         \left(C_L\right)\I{_N_M} \left(A\I{^L_J}\right)\I{_I} \epsilon\I{^I^J^M^N} =  T \I{^\mu_\nu_\rho} \omega \I{_\mu^\rho^\nu}.
     \end{equation}
     \\
     \item TERM 4.\\     
     $$ 2\left(C_I\right)\I{_N_L} \left(B^L\right)\I{_J_M}  \epsilon\I{^I^J^M^N} 
 = \dfrac{1}{4} T \I{_\mu_\nu_I}  e^\mu_R e^\nu_S \epsilon \I{^R^S_N_L} T \I{_\rho_\sigma^L}  e^\rho_J e^\sigma_M \epsilon\I{^I^J^M^N} $$
\begin{equation}\label{eq:A.4}
         = -\dfrac{1}{4} T \I{_\mu_\nu_I} T \I{_\rho_\sigma_L}  e^\mu_R e^\nu_S e^\rho_J e^\sigma_M \epsilon \I{^R^S^L_N} \epsilon\I{^I^J^M^N}
     \end{equation}
 We use the following formula for contracting the Levi-Civita symbol:
     $$\epsilon \I{^R^S^L_N} \epsilon\I{^I^J^M^N} = - \big( \eta^{RI} \eta^{SJ} \eta^{LM} + \eta^{RJ} \eta^{SM} \eta^{LI} + \eta^{RM} \eta^{SI} \eta^{LJ} $$
     \begin{equation}\label{eq:A.5}
     - \eta^{RI} \eta^{SM} \eta^{LJ} - \eta^{RJ} \eta^{LM} \eta^{SI} - \eta^{SJ} \eta^{RM} \eta^{IL}\big)
 \end{equation}
 and substitute it in \eqref{eq:A.4}:
 $$ 2\left(C_I\right)\I{_N_L} \left(B^L\right)\I{_J_M}  \epsilon\I{^I^J^M^N} = -\dfrac{1}{4} T \I{_\mu_\nu_I} T \I{_\rho_\sigma_L} \Big[ g^{\rho \nu} e^{\mu I} e^{\sigma L} + g^{\sigma \nu} g^{\mu \rho} \eta^{IL} $$
 $$+ g^{\sigma \mu} e^{\nu I} e^{\rho L}  - g^{\sigma \nu} e^{\mu I} e^{\rho L} - g^{\rho \mu} e^{\nu I} e^{\sigma L} - g^{\rho \nu} g^{\mu \sigma} \eta^{I L}\Big]. $$
 Hence
 \begin{equation}\label{eq:A.6}
     2\left(C_I\right)\I{_N_L} \left(B^L\right)\I{_J_M}  \epsilon\I{^I^J^M^N} = - T \I{_\mu_\nu^\mu} T \I{_\rho^\nu^\rho} + \dfrac{1}{2} T \I{_\mu_\nu_\rho} T \I{^\mu^\nu^\rho}.
 \end{equation} \\
 \item TERM 5.  \\
 $$2\left(C_I\right)\I{_N_L} \left(A \I{^L_J}\right)_M \epsilon\I{^I^J^M^N} = - \dfrac{1}{2}T \I{_\mu_\nu_I} \omega \I{_\rho^L_J}  e^\mu_R e^\nu_S e^\rho_M \epsilon \I{^R^S^L_N} \epsilon\I{^I^J^M^N}$$
 using again the formula \eqref{eq:A.5} we finally have 
  \begin{equation}\label{eq:A.7}
      2\left(C_I\right)\I{_N_L} \left(A \I{^L_J}\right)_M \epsilon\I{^I^J^M^N} = T \I{_\mu_\nu^\mu} \omega \I{_\rho^\rho^\nu}  + T \I{_\mu^\nu_\rho} \omega \I{_\nu^\rho^\mu}.
  \end{equation}
    \end{itemize}
    \end{appendices}
  	\bibliographystyle{unsrtnat}
  \bibliography{biblioHodge}
  \end{document}